\documentclass{jfm}

\usepackage{graphicx}
\usepackage{natbib}
\usepackage{hyperref}
\hypersetup{
    colorlinks = true,
    urlcolor   = blue,
    citecolor  = black,
}

\newcommand{\RomanNumeralCaps}[1]
\linenumbers

\usepackage{booktabs, multirow} 
\usepackage{soul}
\usepackage{xcolor,colortbl} 
\usepackage{changepage,threeparttable} 

\usepackage{amsmath}
\usepackage{physics}
\usepackage{gensymb}

\usepackage[utf8]{inputenc}
\usepackage[T1]{fontenc}

\usepackage{xcolor} 

\title{Non-equilibrium formulation for inertial particles in turbulent swirling flows}

\author{
B.L.~Espa\~nol\aff{1,2,3},
M. Obligado\aff{4} \corresp{\email{martin.obligado@centralelille.fr}},
J. Peinke\aff{5},
M. Noseda\aff{1,2,3},
P.J. Cobelli\aff{1,2,3},
\and P.D. Mininni\aff{1,2,3}
\corresp{\email{mininni@df.uba.ar}}}

\affiliation{
\aff{1}Universidad de Buenos Aires (UBA), Facultad de Ciencias Exactas y Naturales, Departamento de Física, Ciudad Universitaria, 1428 Buenos Aires, Argentina
\aff{2}CONICET-UBA, Instituto de Física Interdisciplinaria y Aplicada (INFINA), Ciudad Universitaria, 1428 Buenos Aires, Argentina
\aff{3}CNRS-CONICET-UBA, Institut Franco-Argentin de Dynamique des Fluides pour l’Environnement (IFADyFE), IRL2027, Ciudad Universitaria, 1428 Buenos Aires, Argentina
\aff{4}Univ. Lille, CNRS, ONERA, Arts et Metiers Institute of Technology, Centrale Lille, UMR 9014-LMFL-Laboratoire de M\'ecanique des Fluides de Lille - Kamp\'e de F\'eriet, F-59000 Lille, France
\aff{5}Institute of Physics, University of Oldenburg, K\"upkersweg 70, 26129 Oldenburg, Germany
}

\begin{document}
\maketitle

\begin{abstract}

We study the dynamics of inertial particles in turbulence using datasets obtained from both direct numerical simulations and laboratory experiments of turbulent swirling flows. By analyzing time series of particle velocity increments at different scales, we show that their evolution is consistent with a Markov process across the inertial range. This Markovian character enables a coarse-grained description of particle dynamics through a Fokker--Planck equation, from which we can extract drift and diffusion coefficients directly from the data. The inferred coefficients reveal scale-dependent relaxation and noise amplitudes, indicative of inertial filtering and intermittency effects. Beyond the kinematic description, we analyze the thermodynamic properties of particle trajectories by computing the trajectory-dependent entropy production. We show that the statistics of entropy fluctuations satisfy both the Integral Fluctuation Theorem and, under certain conditions, the Detailed Fluctuation Theorem. These results establish a quantitative bridge between stochastic thermodynamics and particle-laden flows, and open the door to modeling turbulent transport using effective stochastic theories constrained by data and physical consistency.

\end{abstract}

\section{Introduction}

The dynamics of inertial particles in turbulent flows are a problem of scientific and practical interest, with applications ranging from industrial transport processes to environmental and atmospheric sciences \citep{Balkovsky_2001, Calzavarini_2008, Homann_2010, Fiabane_2012, Obligado_2014, Xu_2014, Ichihara_2023}. As a specific example, understanding how particles disperse, cluster, and interact with the underlying flow is crucial for improving models of pollutant transport, aerosol dispersion, and cloud microphysics \citep{Shaw_2003}. However, in many of these contexts, a complete description of the three-dimensional turbulent flow is either inaccessible or prohibitively expensive to obtain. Furthermore, the initial conditions of the particles are often poorly constrained. As a result, deterministic forecasting of particle trajectories can in many cases be less meaningful or even impossible, while a statistical description in those cases can, in principle, provide a more robust and insightful alternative \citep{Elghobashi_1994, Falkovich_2001, Drivas_2017}.

In practice, many phenomenological stochastic models were developed in recent years to derive effective equations for particle laden flows and to validate them. As an example, Langevin equations for the particles' position, velocity, and drag force were presented in \citet{Lattanzi_2020}. Stochastic processes were also considered to describe tracers' velocity and acceleration \citep{Drivas_2017, Viggiano_2020}, in many cases with a special focus on the effect of the flow multi-fractality \citep{Peixoto_2023}. Stochastic models for the probability density function (PDF) of dense particle laden flows were also considered by \cite{Innocenti_2019}, and neural networks were used to model drift and diffusion coefficients in stochastic differential equations for particles' velocity fluctuations \citep{Williams_2022}.

A proper statistical description for this problem requires knowledge of the PDFs of particle positions and velocities in a turbulent flow, for single particles, or for multiple particles. While direct numerical simulations (DNSs) of particle-laden turbulence have become widespread and can offer detailed insights \citep{Calzavarini_2008, Homann_2010, Bec_2024}, simulations are typically carried out for a specific flow realization and specific particles' initial conditions. And while PDFs can be computed from them, practical applications in most cases require ensemble-averaged information that captures the essential stochastic behaviour of the particles across scales. A promising route in this direction is the formulation of an effective Fokker--Planck equation that describes the evolution of the PDF of the turbulent system in time or in scale space. This has been successfully done for the Eulerian velocity field of a single-phase turbulent flow in \citet{Friedrich_1997}, \citet{Reinke_2018}, and \citet{Peinke_2019} (see also a more recent study following similar ideas by \citet{Yao_2023}, using instead phenomenological arguments and without a Fokker--Planck formulation). It has also been done for the Lagrangian velocity field formulation of isotropic and homogeneous turbulent flows in \citet{Fuchs_2022}. In the case of particle-laden flows in which the particles have their own inertia or finite radius, if the particles dynamics can be shown to be Markovian at an appropriate level of coarse-graining, then a Fokker--Planck formulation can provide a powerful tool to extract drift and diffusion coefficients, which can in turn encapsulate the essential physics of flow-particle interactions across scales. It is important to remark in this context that these formulations are based on velocity increments, thus properly capturing the multiscale nature of turbulent cascades.

Besides capturing the joint multiscale statistics of increments,  such a Fokker--Planck framework can also provide a gateway into the stochastic thermodynamics of inertial particles in turbulence. For non-equilibrium thermodynamics, the concept of a ``trajectory'' of velocity increments evolving from large to small scales is important. From it, through the drift and diffusion terms in the Fokker--Planck equation, one can define a cascade trajectory-dependent information entropy production, and assess the system's consistency with fundamental fluctuation theorems, such as the Integral Fluctuation Theorem (IFT) and the Detailed Fluctuation Theorem (DFT). These results would not only reinforce the statistical description of turbulence, but also provide a thermodynamic interpretation of turbulent transport processes.

In this work, we investigate whether a Fokker--Planck formulation can be used to study inertial particles in turbulent flows in general cases that include flows with inhomogeneities, anisotropies, and a mean large-scale flow, by testing the Markovian character of the particle dynamics across a range of Stokes numbers. Compared with a previous Lagrangian study by \cite{Fuchs_2022b}, here we focus on inertial particles in both simulations and experiments, thus considering both simple particle models and real particles with finite-size effects, and also flows with a large-scale coherent component superimposed to the turbulence. 
To this end we combine DNSs of inertial particles using different particle models, and laboratory experiments with neutrally buoyant spherical particles. The numerical simulations are performed in a Taylor-Green flow, a periodic flow that sustains turbulence with mean flow structures, inhomogeneity, and anisotropy. In parallel, we analyze trajectories from particle tracking velocimetry (PTV) measurements of spherical particles seeded in a von Kármán water experiment, which, while differing in boundary conditions and forcing mechanisms, shares several Eulerian and Lagrangian features with the Taylor-Green flow \citep{Mordant_2004, Volk_2008, Angriman_2022b, Espanol_2025}. This dual approach allows us to consider different particles' simplified models as well as real particles in the laboratory, to validate methods against numerical and experimental data, and to explore the robustness of the Fokker--Planck description in physically distinct but dynamically similar flow environments, confirming the Markovian behaviour of different particles over the inertial range of turbulent flows, and the validity of IFT and DFT under different regimes, parameters, and physical scenarios.

\section{Methods}

\subsection{The Friedrich--Peinke approach to turbulence}\label{sec:fp-aproach}

Our objective is to obtain a Fokker--Planck description of particles' velocity fluctuations grounded in the physics of the turbulent cascade. To this end we will treat these fluctuations not directly as a function of time, but through velocity increments across decreasing time lags. Velocity increments are the fundamental building blocks of turbulence theory, capturing how energy is distributed at different scales; by computing them along particle trajectories we will construct ``cascade trajectories'' in time-scale space, that mirror the classical cascade in scale space. Afterwards, we will coarse-grain these increments using the smallest possible time step such that the process becomes Markovian. This is essential in stochastic thermodynamics---where Markovianity is required to define a Fokker--Planck equation---but it also has a clear physical meaning in turbulence: near the fluid viscous (or particle Stokes) time, dissipation and drag are expected to regularize the dynamics and introduce memory \citep{Daitche2015}, so increments at too short time lags can be expected to be more correlated. Once the appropriate coarse-graining is identified, we will test whether the resulting cascade trajectories indeed satisfy a Fokker--Planck equation. When they do, the Fokker--Planck description will provide a stochastic analogue of the turbulent cascade itself: fluctuations flow from large to small time increments, just as turbulent eddies transfer energy downscale. This explains why signs in the following Fokker--Planck formulation will seem reversed---the evolution variable will be the decreasing time scale, not the increasing physical time---as the stochastic process will march forward as the cascade progresses towards shorter times.

Let's then consider a time series of a velocity component $u(t)$ (which can be any Cartesian component of a particle's velocity, or of the fluid velocity at some point, as a function of time). In particular, here we consider velocity increments over Lagrangian trajectories defined as $u_\tau(t) = u(t + \tau) - u(t)$, where $\tau$ belongs to a sequence of time lags $\tau_0 > \tau_1 > \dots > \tau_n$ equally spaced by an interval $\Delta\tau$, such that $\tau_j - \tau_{j+1} = \Delta\tau$ (note that $\tau_j$ decreases with increasing $j$). If a coarse-grained version of $u_\tau(t)$ for time increments larger than some Einstein--Markov time $\Delta_\text{EM}$ is Markovian, i.e., if $\Delta\tau$ is sufficiently large for $u_\tau$ to satisfy the Markov property in $\tau$, then the $\tau$-evolution of the stochastic variable $u_\tau$ as a function of the scale $\tau$ can be described by a diffusion process \citep{Luck_2006},
\begin{equation}
	-\dd u_\tau = \,D^{(1)}(u_\tau, \tau)\,\dd\tau
	+ \sqrt{2D^{(2)}(u_\tau, \tau)}~\dd W_\tau,
	\label{eq:Langevin}
\end{equation}
where $W_\tau$ is a Wiener process with zero mean and independent increments, characterized by $\ev{\dd W_\tau} = 0$ and $\ev{\dd W_\tau^2} = \dd\tau$.
The functions $D^{(1)}(u_\tau, \tau)$ and $D^{(2)}(u_\tau, \tau)$ correspond to the drift and diffusion coefficients, respectively. These functions are, in general, nontrivial, and depend explicitly on both the scale $\tau$ and on the value of the increment $u_\tau$ at that scale.
This equation is equivalent, in the It\^o sense, to the Fokker--Planck representation for the PDF of $u_\tau$, single-conditioned to $u_{\tau_0}$, given by \citet{Friedrich_1997}
\begin{equation} \label{eq:fp}
	-\partial_\tau\,p(u_\tau | u_{\tau_0}) =
	- \partial_{u_\tau}\big[D^{(1)}(u_\tau,\tau)\,p(u_\tau | u_{\tau_0})\big]
	+ \partial_{u_\tau}^{2}\,\big[D^{(2)}(u_\tau | \tau)\,p(u_\tau| u_{\tau_0})\big].
\end{equation}

To estimate the Fokker--Planck equation that best reproduces the data we follow the procedure described in \citet{Peinke_2019} and \citet{Fuchs_2022}.
Briefly, once the Einstein--Markov time has been identified, the drift and diffusion coefficients can be expressed as
\begin{equation}
	D^{(k)}(u_\tau, \tau) = \lim_{\Delta\tau \to 0} \frac{M^{(k)}(u_\tau, \tau, \Delta\tau)}{k!\, \Delta\tau},
	\quad k = 1, 2, \quad \tau-\tau^\prime=\Delta\tau,
\end{equation}
where the conditional moments $M^{(k)}$ are defined as
\begin{equation}
	M^{(k)}(u_\tau, \tau, \Delta\tau) = \int_{-\infty}^\infty \left(u_{\tau^\prime} - u_\tau\right)^k p(u_{\tau^\prime}|u_\tau)\, \dd u_{\tau^\prime},
\end{equation}
which can be estimated directly from numerical or experimental data using the optimization procedure proposed in \citet{Peinke_2019}. This procedure was carried out using the open-source Python package \texttt{turbfpe} (\url{https://github.com/bersp/turbfpe}), which builds upon a previous MATLAB implementation called \texttt{OPEN\_FPE\_IFT} \citep{Fuchs_2022b}.

\subsection{Numerical dataset}

The numerical datasets stem from DNSs of particle-laden flows. For the Taylor-Green (TG) dataset we performed a DNS of the incompressible Navier-Stokes equation using an external forcing $\mathbf{F}$ to sustain a turbulent regime. The equations are solved in a 3D cubic, $(2 \pi L_0)^3$-periodic domain where $L_0$ is a unit length, using a parallel pseudospectral method with the GHOST code \citep{Mininni_2011c, Rosenberg_2020}, with a spatial resolution of $N^3$ = $768^3$ grid points. The forcing $\mathbf{F}=(F_x,F_y,F_z)$ is given by
\begin{equation}
	F_x = F_0 \sin x \cos y \cos z, \ \ \ F_y = -F_0 \cos x \sin y \cos z, \ \ \ F_z = 0 .
\end{equation}
This forcing results in two counter-rotating large-scale toroidal vortices in a $(\pi L_0)^3$ subvolume, with a secondary poloidal circulation. The flow in each subvolume of size $(\pi L_0)^3$, inside the entire $(2 \pi L_0)^3$-periodic volume, resembles the flow used in the laboratory experiments \citep{Angriman_2022b, Espanol_2025}. In the turbulent steady state of these simulations, the turbulent Reynolds number is $R_\lambda \approx 350$, and the ratio of the mean flow velocity to the turbulent fluctuations is $\ev{U}/u' \approx 0.59$, where $\ev{U}$ denotes the root mean square space-time averaged value of the Eulerian velocity field, and $u'$ the root mean square turbulent fluctuations. 

When the turbulent steady state in the simulations is reached, we evolve in time the position and velocity of sets of $10^6$ particles, each set with different Stokes numbers and with different equations of motion.
In all sets, particle trajectories were integrated for at least three turbulence integral correlation times. For small neutrally buoyant particles we solve the Maxey-Riley-Gatignol equation \citep{Maxey_1983, Gatignol_1983}, neglecting the Basset-Boussinesq term (we explicitly verified that preserving this term yields similar results in the analysis below),
\begin{equation}
	\frac{d{\bf v}}{dt} = \frac{1}{\tau_p} \left[ {\bf u}({\bf x},t) - {\bf v}(t) \right] + \frac{\textrm{D}}{\textrm{D}t} {\bf u}({\bf x},t) ,
    \label{eq:MR}
\end{equation}
where $\tau_p = a^2/(3 \nu)$ is the particle response time, $a$ is the particle radius, ${\bf v}(t)$ is the particle velocity, and ${\bf u}({\bf x},t)$ is the fluid velocity at the particle position.
For big particles, linear drag and added mass forces do not provide a good approximation to their dynamics \citep{Reijtenbagh_2023}.
We therefore consider instead a simplified model of small heavy particles only with nonlinear drag \citep{Wang_1993, Stout_1995} and with negligible settling velocity (note that we also neglect added mass forces \citep{Brandt_2022}),
\begin{equation}
	\frac{d{\bf v}}{dt} = \frac{1+0.15 \textrm{Re}_p^{0.687}}{\tau_p} \left[ {\bf u}({\bf x},t) - {\bf v}(t) \right] ,
    \label{eq:NLD}
\end{equation}
where $\textrm{Re}_p = (18 \tau_p/\nu)^{1/2} |{\bf u}({\bf x},t) - {\bf v}(t)|$ is the instantaneous Reynolds number of the particle. Equation (\ref{eq:MR}) is valid for small particles with $\textrm{Re}_p < 1$, while Eq.~(\ref{eq:MR}) is expected to hold, at least, for $1 < \textrm{Re}_p < 40$ \citep{Wang_1993}.

To have some parameters similar to those in the experimental datasets, we performed four simulations with particles with different $\tau_p$ and different Stokes numbers, $\textrm{St}=\tau_p/\tau_\eta$, where $\tau_\eta$ is the Kolmogorov dissipation time of the flow. We solved two sets of particles with the Maxey-Riley (MR) equation with $\textrm{St}=0.76$ and $3.2$, labelled respectively as TG-0.76 and TG-3.2 (MR). And two sets of particles with nonlinear drag (NLD), with $\textrm{St}=3.2$ and $8.9$, labeled respectively as TG-3.2 and TG-8.89. Note that we integrated two different sets of particles with $\textrm{St}=3.2$, in one case using the Maxey-Riley equation and in the other case using nonlinear drag, to compare, for particles with moderate $\textrm{St}$, the results obtained from both models against the experimental data (see table \ref{tab:parameters}).

It is also important to stress that our particle models are intentionally simple: they are selected only to reproduce variations in the particle response time \citep{Brandt_2022, Bec_2024} and in $\mathrm{Re}_p$ comparable to those in the experiment. The full dynamics of particles in turbulent flows---including deviations from tracer behaviour, preferential clustering, sensitivity to local flow deformation, and two-way coupling---depends on additional parameters \citep{Calzavarini_2009, Fiabane_2013}. Moreover, the experiment will use neutrally buoyant particles, so at large $\mathrm{St}$ the model in Eq.~(\ref{eq:NLD}) will not capture several relevant features. The experiments should therefore be viewed as a complimentary validation of the non-equilibrium formulation: they probe particles with similar response times and $\mathrm{Re}_p$, but with other---significantly different---physical properties.

\begin{table}
\caption{Parameters of the simulations and experiments. In the datasets, the labels indicate Taylor-Green (TG) or von Kármán (VK) flows, followed by the Stokes number of the particles, $\textrm{St}$. Model indicates whether the Maxey-Riley (MR) or nonlinear drag (NLD) equations were solved in the simulations, $\Delta_{EM,x}/\tau_\eta$ is the Einstein--Markov time for the $x$ component of the particles' velocity in units of the Kolmogorov dissipation time, $\Delta_{EM,z}/\tau_\eta$ is the same for the $z$ velocity component, and $\langle \textrm{Re}_p \rangle$ is the averaged particles' Reynolds number.}
	\centering
	\label{tab:parameters}
	\begin{tabular}{ccccccccc}
		Dataset & St & Model & $\Delta_{\text{EM},x}/\tau_\eta$ & $\Delta_{\text{EM},z}/\tau_\eta$
		& $\langle \text{Re}_p \rangle$ \\
		\hline
		TG-0.76      & $0.76$ & MR  & $7.44\pm0.37$ & $6.33\pm0.37$ & 0.2 \\
		TG-3.2 (MR) & $3.2$ & MR  & $7.44\pm0.37$ & $6.33\pm0.37$ & 2.5 \\
		TG-3.2      & $3.2$ & NLD & $13.0\pm0.37$ & $9.3\pm0.37$ & 5.8 \\
		TG-8.89      & $8.89$ & NLD & $18.6\pm0.37$ & $11.9\pm0.37$ & 14.9 \\
		VK-0.7       & $0.7\pm 0.2$ & --  & $6.50\pm0.13$ & $6.0\pm0.13$  & $<1$ \\
		VK-2.0       & $2.0\pm 0.5$ & --  & $8.75\pm0.25$ & $7.25\pm0.25$ & $4.3$ \\
		VK-8.0       & $8.0\pm2.0$ & --  & $10.0\pm0.25$ & $9.25\pm0.25$ & $8.7$ \\
	\end{tabular}
\end{table}

\subsection{Experimental dataset}

The von Kármán (VK) experimental setup consists of a transparent-wall tank with an interior volume of $(20 \times 20 \times 50)$\,cm$^3$, featuring two facing disks of diameter $D = 19 \ \rm{cm}$, separated by a distance of $H = 20 \ \rm{cm}$. Each disk is fitted with eight radially oriented straight blades, each having a height and width of 1 cm; limited in length so they do not intersect at the center \citep{Angriman_2020, Angriman_2022b}. These two disks with blades work as impellers, driven by two independent brushless rotary motors (Yaskawa SMGV-20D3A61, 1.8 kW) controlled by servo-controllers (Yaskawa SGDV-8R4D01A).
On the back of the impellers, two coils connected to a chiller allow heat removal, keeping the temperature of the working fluid at 25 $\degree$C. Distilled water from a double-pass reverse osmosis system is used as the working fluid. A forcing frequency of $f_0 = 1/T_0 = 50 \, \rm{rpm}$ was used for the rotary motors in all experiments, with the two impellers counter-rotating. The forcing generates two counter-rotating structures in the working fluid, two poloidal secondary circulations, and a strong shear layer on the mid-plane of the experiment, which shares similarities with the TG flow used in the numerical simulations \citep{Angriman_2022b, Espanol_2025}. The generated flow is inhomogeneous and anisotropic, with large-scale structures parallel to the disks' planes (horizontal) that have a larger correlation length than their axial (vertical) counterparts. As a reference for a comparison with the numerical flow, in the turbulent steady state of this flow $R_\lambda \approx 240$, and the ratio of the mean flow velocity to the turbulent fluctuations is $\ev{U} /u' \approx 0.61$. Albeit there is a small difference in $R_\lambda$ between experiments and simulations, the turbulent fluctuations are comparable, and comparisons between these flows with similar differences in $R_\lambda$ have been carried on before \citep{Angriman_2022b}.

We seeded this flow with neutrally buoyant polyethylene microspheres (density of $1 \pm 0.01 \, \rm{g} \, \rm{cm}^{-3}$) from Cospheric. Three different particles' diameters were considered successively, with mean radii of $a=(137.5 \pm 13)$ $\mu$m, $(275 \pm 18.75)$ $\mu$m, and $(462.5 \pm 37)$ $\mu$m, which resulted respectively in Stokes numbers of $\textrm{St} = 0.7 \pm 0.2$, $2.0 \pm 0.5$, and $8.0 \pm 2.0$ (see table \ref{tab:parameters}). As a reference, the Kolmogorov scale in the experiment is $\eta = (93\pm2)$ $\mu$m.
Two $(25 \times 25)$~cm$^2$ LED panels (1880 lm, 22 W each) were used to illuminate the particles. The instantaneous position of the particles was recorded by two high-speed Photron FASTCAM SA3 cameras ($1024 \times 1024$ px$^{2}$, 12 bit), positioned to capture their dynamics from two mutually perpendicular projections sharing the vertical coordinate. The cameras were located 2.5~m from the center of the tank, and fitted with 250~mm focal length objective lenses to minimize parallax errors. 
A stereo calibration was performed by placing a checkerboard target of known spacing inside the measurement region and imaging it simultaneously with both cameras, whose optical axes were rigidly aligned with the transparent faces of the tank, with one view folded by a front-surface mirror. Using the shadow-PTV procedure of Huck (2017), we verified that residual perspective distortions remained below the particle diameter, and we reconstructed the 3D position of each particle by averaging the redundant coordinate obtained from the two calibrated projections.
From the particle' positions, trajectories and velocities were obtained through three-dimensional Particle Tracking Velocimetry (3D-PTV) (see \citet{Angriman_2020, Angriman_2022b} for more details). 
This setup allows for tracking particles within a cubic region centered on the tank and spanning a volume of $(16 \times 16 \times 16) \ \rm{cm}^3$.

In the following, we label the experimental datasets obtained with this technique as VK followed by the Stokes number of the particles considered. The VK-0.7 dataset contains $\approx 30,000$ trajectories, while the VK-2.0 and VK-8.0 datasets each contain $\approx 5,000$ trajectories. As particles' trajectories end when they exit the measurement volume, their lengths have a broad distribution. As a reference, all trajectories exceed half a turbulence integral correlation time, and at least $1,500$ in each dataset persist for more than one correlation time.
The mean distance between particles in the measurement volume was $0.014 \pm 0.006$ m, $0.0215 \pm 0.009$ m, and $0.03 \pm 0.01$ m, respectively for the three sets of particles with increasing radius. This results in all cases in volume fractions $\Phi_p = V_p/V_f$ (i.e., the ratio of the total volume occupied by particles to the fluid volume) of $\mathcal{O}(10^{-6})$, in the upper limit to still consider the particles as one-way coupled \citep{Elghobashi_1994}.

\section{Results}

\begin{figure}
	\centering
	\includegraphics[width=\linewidth]{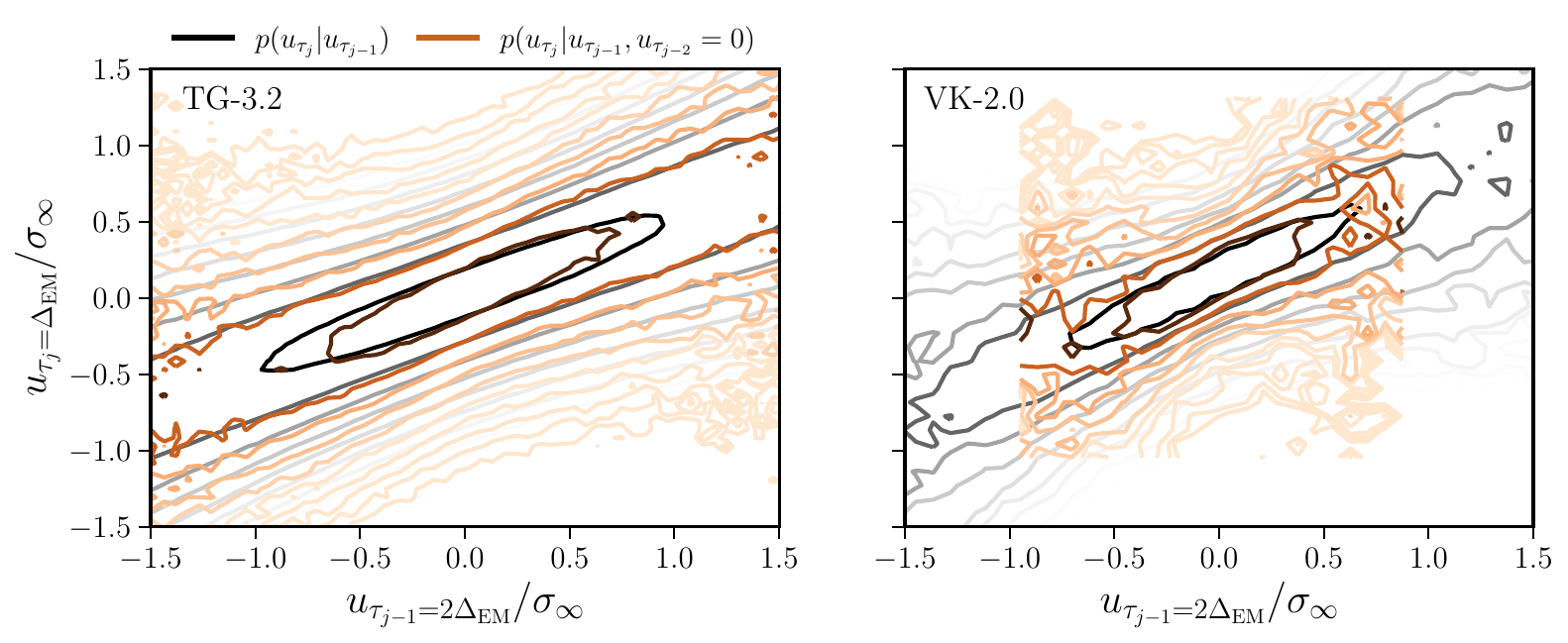}
	\caption{Contour plots comparing the single-conditioned probability (black contours) and the double-conditioned probability (red contours) of velocity increments of particles at three time scales, $\tau_2 < \tau_1 < \tau_0$. As a representative example, the time scales are chosen relative to the Einstein--Markov time of each dataset: $\tau_2 = \Delta_\text{EM}$, $\tau_1 = 2\tau_2$, and $\tau_0 = 3\tau_2$. The left panel shows results for the TG-3.2 dataset, and the right panel for VK-2.0. Velocity increments are normalized using $\sigma_\infty = 2\sigma$, where $\sigma$ is the standard deviation of the increments. Note that the choice $u_{\tau_0} = 0$ is arbitrary; changing this value shifts the center of the conditioned distributions but still yields comparable results.}
	\label{fig:2dpdf}
\end{figure}

Given the anisotropy of the TG and VK datasets, we will consider time series $u(t)$ of the particles' velocities separating two Cartesian components: the $x$ component which is perpendicular to the axis of symmetry of the impellers in VK and of the large-scale vortices in TG, $u(t) = v_x(t)$, and the $z$ component which is parallel to this axis, $u(t) = v_z(t)$. We verified that repeating the analysis using the $y$ velocity component yields similar results as for $v_x(t)$. Both the TG and VK flows have a large-scale mean flow  with two toroidal counter-rotating structures (a ``shearing mode''), and a poloidal circulation (a ``poloidal mode'') \citep{Green_1937, Berning_2023, Espanol_2025}. The poloidal mode has approximately half the correlation length of the shearing mode, and as a result, the correlation time of $v_z$ is smaller than the correlation time of $v_x$. This implies that the inertial range visible in the spectrum of $v_z$ is shorter than that of $v_x$ and $v_y$, further pointing to the need to separate the analysis of these two velocity components.

For a velocity component, a stochastic process on the velocity increment $u_\tau = u(t+\tau) - u(t)$ is said to be Markovian in the coarse-grained scale $\tau$ if the conditional probability distribution satisfies
\begin{equation}
    p(u_{\tau_j} \mid u_{\tau_{j-1}}) = p(u_{\tau_j} \mid u_{\tau_{j-1}}, u_{\tau_{j-2}}, \ldots),
\end{equation}
where the time increments (or ``scales'') $\tau_j, \tau_{j-1}, \tau_{j-2}, \ldots$ are separated by some fixed increment $\Delta$, with $\tau_j = \tau_{j-1} - \Delta$ such that $\tau_{j} < \tau_{j-1}$, and all $\tau_j$ lie within the inertial range.
Note that saying that the process is Markovian is equivalent to saying that the statistics of a fine-scale coarse-grained velocity increment conditioned on coarser representations depend only on the immediately preceding increment, and not on the full history at larger scales (or larger time increments).
While this is consistent with the turbulence picture of the direct energy cascade in the inertial range \citep{Reinke_2018, Peinke_2019}, where information is transferred locally in scale from larger to smaller scales, it is not trivial that the velocity fluctuations of a particle with its own inertia submerged in a turbulent flow must satisfy this condition.

In practice, for discrete signals, the validity of the Markov property is typically assessed by verifying the simplified condition
\begin{equation}
	p(u_{\tau_{j}} \mid u_{\tau_{j-1}}) = p(u_{\tau_j} \mid u_{\tau_{j-1}}, u_{\tau_{j-2}}) \label{eq:markov_prop}
\end{equation}
for different values of $\tau_j$ \citep{Renner_2001, Peinke_2019, Fuchs_2022}.
Figure \ref{fig:2dpdf} shows examples of these two conditional probability distributions, computed from the TG-3.2 and VK-2.0 datasets, each evaluated at a scale separation corresponding to its respective Einstein--Markov coherence time scale $\Delta_{\text{EM}}$.
This is the smallest time increment at which the system exhibits Markovian behaviour; that is, it is the finest time coarse-graining for which Eq.~\eqref{eq:markov_prop} is qualitatively satisfied.
For time scales smaller than $\Delta_\text{EM}$ the system behaves deterministically (and is therefore non-Markovian), whereas for much larger scales it becomes fully decorrelated,
i.e., $p(u_{\tau_j} \mid u_{\tau_{j-1}}) = p(u_{\tau_j})$. This analysis was repeated for all datasets and for different values of $u_{\tau_{j-2}}$.

\begin{figure}
	\centering
	\includegraphics[width=\linewidth]{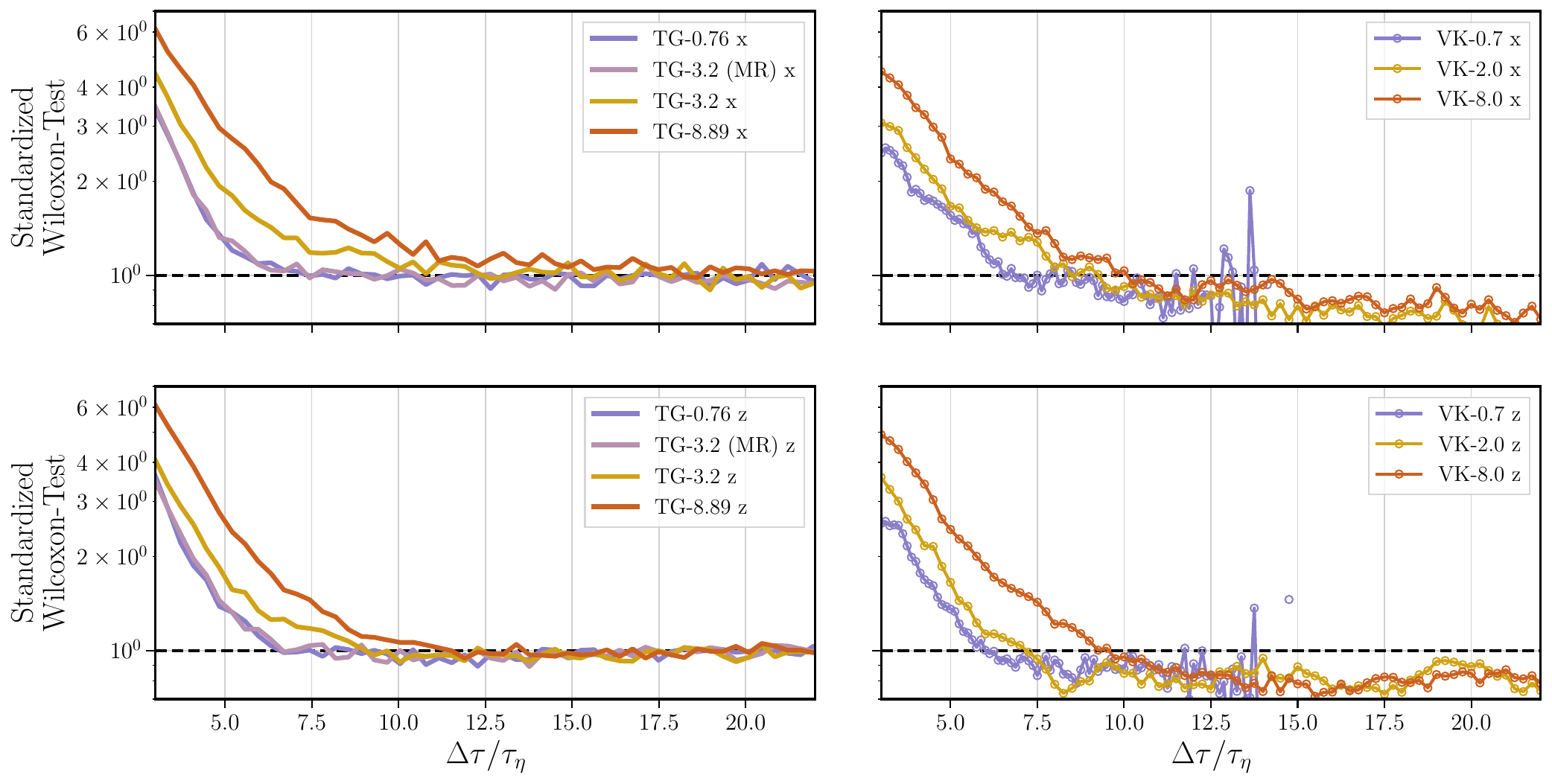}
	\caption{Standardized Wilcoxon test statistic as a function of the time lag $\Delta \tau$ for all datasets analyzed. The time lag is given in units of the dissipation time $\tau_\eta$ of each flow. 
    The left panels correspond to simulations, and the right panels to experimental data. The top row shows results for the $x$ velocity component, while the bottom row corresponds to the $z$ component. The horizontal dashed black line marks the threshold value of $1$, which corresponds to the expected result when the Markovian assumption holds.}
    \label{fig:wilcoxon}
\end{figure}

To systematically assess the similarity between the conditional distributions, and to quantitatively identify the Einstein--Markov scale, we use the Wilcoxon test, which is a parameter-free statistical test to compare two samples, as previously used in \citet{Luck_2006}.
The test evaluates whether two samples come from the same parent distribution by ranking all values together, computing the sum of ranks for each sample, and comparing this sum with its expected value under the null hypothesis of equality of distributions.
Since the comparison involves conditional distributions in two variables, the test is applied at fixed values of the conditioning increment, and the resulting statistics are then averaged to yield a single indicator of Markovianity for that scale.
The resulting standardized statistic of this test takes a value close to $1$ when the two distributions are statistically indistinguishable, i.e., when the Markov property holds, whereas it takes values significantly larger than $1$ as the system deviates from Markovian behaviour.
Figure~\ref{fig:wilcoxon} shows this statistic as a function of the scale increment $\Delta \tau$ for all datasets analyzed, for both the $x$ and $z$ velocity components. The estimated values of $\Delta_\text{EM}$ for each case are reported in Table~\ref{tab:parameters}.
Several comments are in order: 
First, note that in both datasets $\Delta_{\text{EM}}$ monotonically increases with $\textrm{St}$ (or with $\tau_p$ of the particles).
This is consistent with the observations in \citet{Fuchs_2022}, where tracers were seen to be influenced by all scales of the flow, whereas point particles with inertia were found to filter out the fastest time scales and follow trajectories that remain temporally coherent for longer times.
Second, note that the Maxey-Riley equation yields results compatible with the experiment for $\textrm{St} \approx 0.7$, but fails to capture the increase of $\Delta_{\text{EM}}$ for larger values of $\textrm{St}$. 
In particular, for $\textrm{St} = 3.2$, Fig.~\ref{fig:wilcoxon} shows that the NLD model (i.e., the TG-3.2 curve) reproduces an increase in $\Delta_{\text{EM}}$ with increasing $\tau_p$ as in the experiment, while the MR equation with the same $\textrm{St}$ yields tracer-like behaviour. Or, in other words, note how the TG-0.76 and TG-3.2 (MR) curves collapse in Fig.~\ref{fig:wilcoxon}, while the VK-0.7 and VK-2.0 curves display a distinct trend. We verified that considering the Basset-Boussinesq history term in the Maxey-Riley equation, or even Fax\'en corrections, do not improve this behaviour. Instead, the model using nonlinear drag (keeping in mind its limitations) reproduces the experimental trend reasonably well. This is probably associated with the fact that for all particles with $\textrm{St} \ge 2$, also $\langle \textrm{Re}_p \rangle >1$ (see table \ref{tab:parameters}). As a result, in the following we consider only the numerical dataset TG-3.2 when comparing with the laboratory data with similar $\textrm{St}$. Third, note that both in DNSs and in experiments, the $z$ velocity component becomes Markovian faster than the $x$ component. This is consistent with the anisotropic structure of these flows discussed before, in which energy is injected into the shearing mode, and the poloidal mode arises as part of the energy cascade, with a shorter correlation length and time.

\begin{figure}
	\centering
	\includegraphics[width=\linewidth]{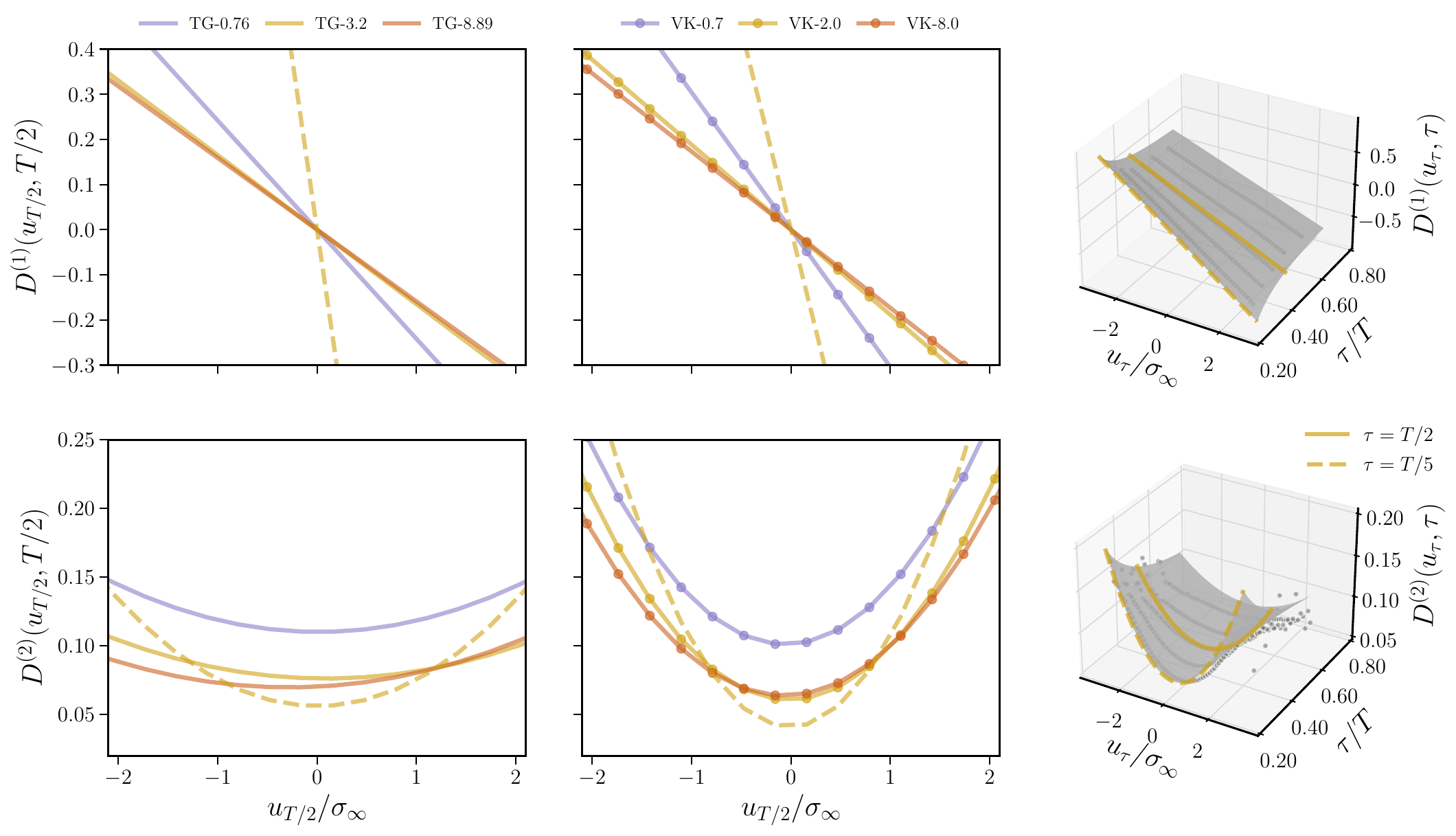}
	\caption{
		Drift coefficients $D^{(1)}(u_\tau, \tau)$ (top panels) and diffusion coefficients $D^{(2)}(u_\tau, \tau)$ (bottom panels) for the $x$ velocity component.
        Solid curves correspond to the coefficients at $\tau = T/2$. The dashed yellow lines indicate the same coefficients for the TG-3.2 and VK-2.0 datasets at $\tau = T/5$; the solid and dashed yellow lines allow for a comparison of the coefficients at different $\tau$. 
        The 3D plots on the right display these coefficients as functions of $u_\tau$ and $\tau$ for the TG-3.2 simulation, used as a representative example (all datasets exhibit similar features). 
		The overlaid solid and dashed yellow curves indicate the cross-section at $\tau = T/2$ and at $\tau = T/5$, respectively.
		In all cases, time lags are normalized by $T$, and increments of the velocity by the asymptotic standard deviation $\sigma_\infty$.
	}
	\label{fig:d12}
\end{figure}

The drift and diffusion coefficients for all datasets can be obtained using the procedure described in Sec.~\ref{sec:fp-aproach}. 
As a representative example, Fig.~\ref{fig:d12} shows the resulting drift and diffusion coefficients for the $x$ velocity component in the TG-3.2 dataset. Figure \ref{fig:d12} also displays $\tau=T/2$ slices of these coefficients for all datasets considered, plus one numerical and one experimental case evaluated at a shorter time lag $\tau=T/5$, where $T$ is the turbulence integral correlation time. 
We discuss first the coefficients at fixed time lag, with $\tau=T/2$. 
The drift coefficients exhibit an approximately linear dependence on $u_\tau$ for all datasets, $D^{(1)}(u_\tau, \tau) = -\gamma(\tau)u_\tau$, suggesting a relaxation mechanism compatible with an Ornstein-Uhlenbeck process for each time lag $\tau$.
Note that a linear dependence of this coefficient with $u_\tau$ for lags larger than the Kolmogorov time is consistent with the expected near-exponential decay of Lagrangian velocity correlations. Recovering this trend directly from the empirical Kramers--Moyal coefficients confirms that our method correctly captures the particles' relaxation for intermediate and large lags.
The parameter $\gamma(\tau)$ then sets the inverse correlation time at that scale, implying that smaller values of $\gamma(\tau)$ correspond to smoother trajectories. This observation supports the interpretation of the drag (controlled by the Stokes number) as an effective low-pass filter of flow fluctuations, with larger St values attenuating rapid fluctuations. Moreover, note that the VK and TG datasets with similar St display comparable behaviour and values of $\gamma$.

The diffusion coefficient, $D^{(2)}(u_\tau, \tau)$, also shown in Fig.~\ref{fig:d12}, plays a different role: it controls the amplitude of the stochastic forcing, with larger values corresponding to stronger noise.  $D^{(2)}(u_\tau, \tau)$ increases significantly for large values of $|u_\tau|$, indicating that large velocity excursions are more likely to be followed by much larger fluctuations at a smaller (faster) scale. This scale-dependent amplification of fluctuations is a key signature of intermittency in turbulence, and underpins the emergence of non-Gaussian statistics in the distribution of $u_\tau$.
Note that as our method does not assume that noise is of the additive type in Eq.~(\ref{eq:Langevin}), the behaviour seen in $D^{(2)}$ reflects genuine nonlinearities of the inertial cascade that cannot be not captured by a linear Langevin equation for Brownian-like motion.
Indeed, the observed features can be interpreted within a Fokker--Planck framework with multiplicative noise, where the scale-dependent drift and diffusion coefficients are respectively given by $D^{(1)} = -\gamma(\tau) u_\tau$ and $D^{(2)} = \alpha(\tau) u_\tau^2 + \beta(\tau)$ (with $\gamma, \alpha, \beta >0$). 
Non-Gaussian contributions to the statistics of velocity fluctuations are associated to the $D^{(2)}$ dependence on $u_\tau$. Neglecting $\beta$ in $D^{(2)}$, the model predicts a log-normal distribution of the increments $u_\tau$, with the intermittency level governed by $\alpha(\tau)$ \citep{Falkovich_2001}. In this context, the value of $\beta(\tau)$ can be considered as a measure of the deviations of the statistics from log-normality.
Focusing again in the cases with a fixed time lag $\tau = T/2$, in both the experiments and the simulations $\beta(\tau)$ decreases with $\textrm{St}$ and has comparable values for particles with similar $\textrm{St}$. 
Overall, the dependence of $\alpha(\tau)$ and $\beta(\tau)$ highlights how inertia acts as a filter of extreme fluctuations: particles with larger St are less responsive to sharp accelerations, and thus experience reduced intermittency, with the particles in the experiment being apparently more sensitive to extreme events than the particles modelled in the DNSs.

In this sense, the compatibility seen at fixed $\tau$ between the $D^{(1)}$ coefficients in experiments and in simulations indicates that simple particle models can capture the drift behaviour of experimental particles' distribution. However, the smaller values of $\alpha(\tau)$ seen in the DNSs indicate that point particles' models filter much more the fluctuations than real particles, reminding us of the models' limitations.

The discussion so far focused on the behaviour of the coefficients in Fig.~\ref{fig:d12} with different $\textrm{St}$ and fixed  $\tau = T/2$. However, the discussed trends extend across the inertial range, as shown in the three-dimensional plots in the same figure, as well as in the dashed curves that show $\tau = T/5 \approx 2.3\Delta_\text{EM}$ for two datasets, illustrating how the coefficients evolve as $\tau$ decreases.
Both the linearity of $D^{(1)}$ and the quadratic dependence of $D^{(2)}$ persist across scales, with their amplitudes decreasing toward the integral scale. For scales larger than the integral time, both $D^{(1)}$ and $D^{(2)}$ tend toward constant values, consistent with Brownian diffusion. 
Conversely, approaching the Kolmogorov scale from above (i.e., for small $\tau$ or for fast time scales) $\gamma(\tau)$ increases, indicating shorter correlation times, while  $D^{(2)}$ grows rapidly increasing its concavity, reflecting the stronger intermittency at those scales. Similar trend can be observed for homogeneous isotropic turbulence. For a comparison of this behaviour at different $\textrm{Re}_\lambda$ in homogeneous isotropic turbulence, see the Appendix.

\begin{figure}
	\centering
	\includegraphics[width=\linewidth]{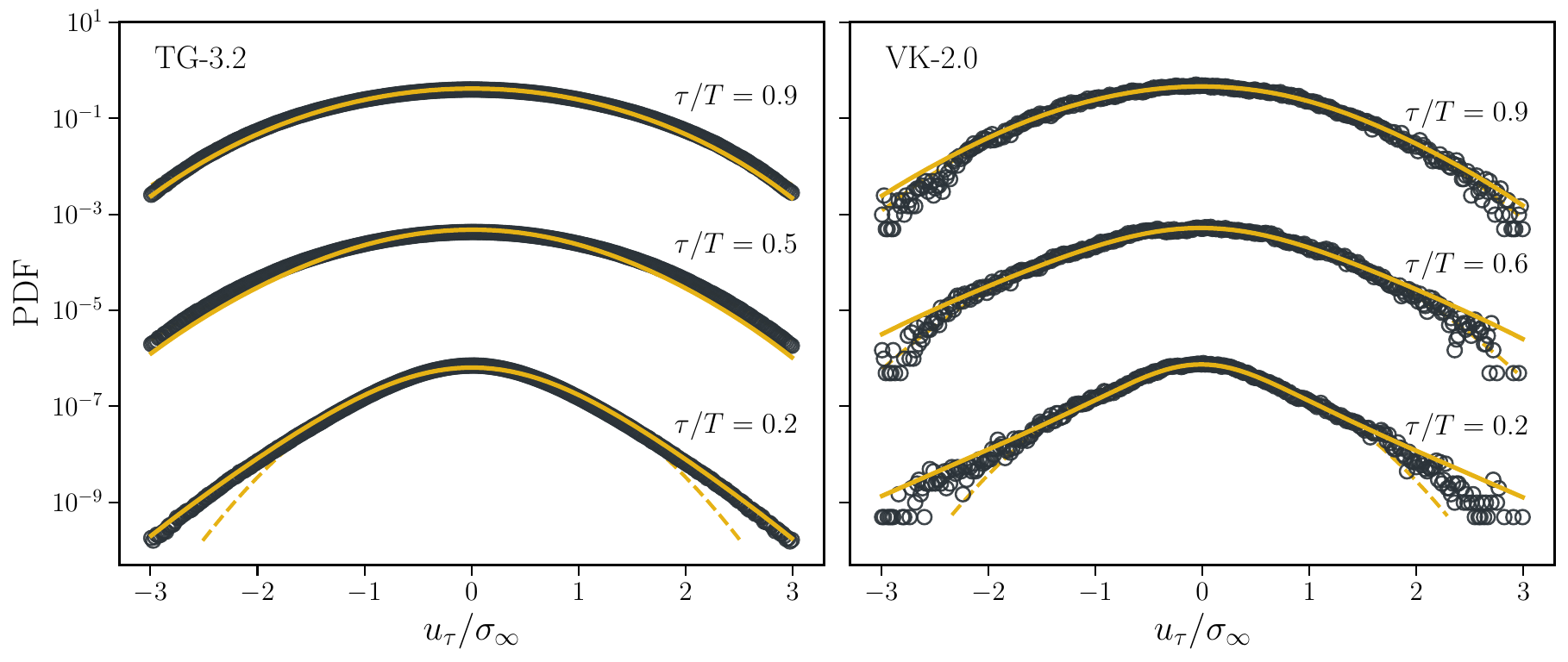}
	\caption{PDFs of velocity increments $u_\tau$ for different time lags (see labels), for the $x$ component of the particles' velocities in the TG-3.2 (left panel) and VK-2.0 datasets (right panel). Black circles show the PDFs computed directly from the data, yellow solid lines show the PDFs obtained from the Fokker--Planck formulation using the short-scale propagator, and dashed yellow lines indicate the Gaussian distribution that best fits the data. Velocity increments are normalized by the asymptotic standard deviation $\sigma_\infty$. For clarity, the curves are vertically shifted by arbitrary offsets.}
	\label{fig:1dpdf}
\end{figure}

Using these Kramers--Moyal coefficients, the conditional probability densities $p(u_\tau|u_{\tau'})$ can be computed directly using the short-time Fokker--Planck propagator (in our case, a ``short-scale'' propagator, as it evolves the probability in scale space),
\begin{equation}
	p_\text{scp}(u_{\tau'}|u_{\tau})
	= \frac{1}{\sqrt{4\pi D^{(2)}(u_\tau, \tau)\Delta \tau}}
	\exp \left[ -\frac{\left(u_{\tau'} - u_\tau - D^{(1)}(u_\tau, \tau)\Delta \tau\right)^2}{4\,D^{(2)}(u_\tau, \tau)\Delta \tau} \right],
\end{equation}
where $\Delta\tau = \tau - \tau'$. This relation connects the statistics at scale $\tau$ to those at nearby (smaller) scales $\tau'$, enabling step-by-step reconstruction of the velocity increment PDFs across the inertial range.
To carry this out, the increment statistics at the integral scale $u_{\tau_0}$ are computed directly from the data, and then iteratively propagated to smaller scales, mirroring the cascade process in which energy injected at large scales is transmitted through the hierarchy of scales.
Figure~\ref{fig:1dpdf} presents examples of the reconstructed density functions of velocity increments at different scales $\tau'$, $p(u_{\tau'}) = \int_{-\infty}^\infty p_\text{scp}(u_{\tau'}|u_{\tau}) p(u_{\tau}) \dd u_\tau $, for the TG-3.2 and VK-2.0 datasets, compared against the actual data.
The reconstructed PDFs align closely with those obtained directly from the data, and reproduce the scale-dependent intermittency observed at small scales. Similar behaviour is observed in all datasets and at all scales within the inertial range.

\begin{figure}
	\centering
	\includegraphics[width=\linewidth]{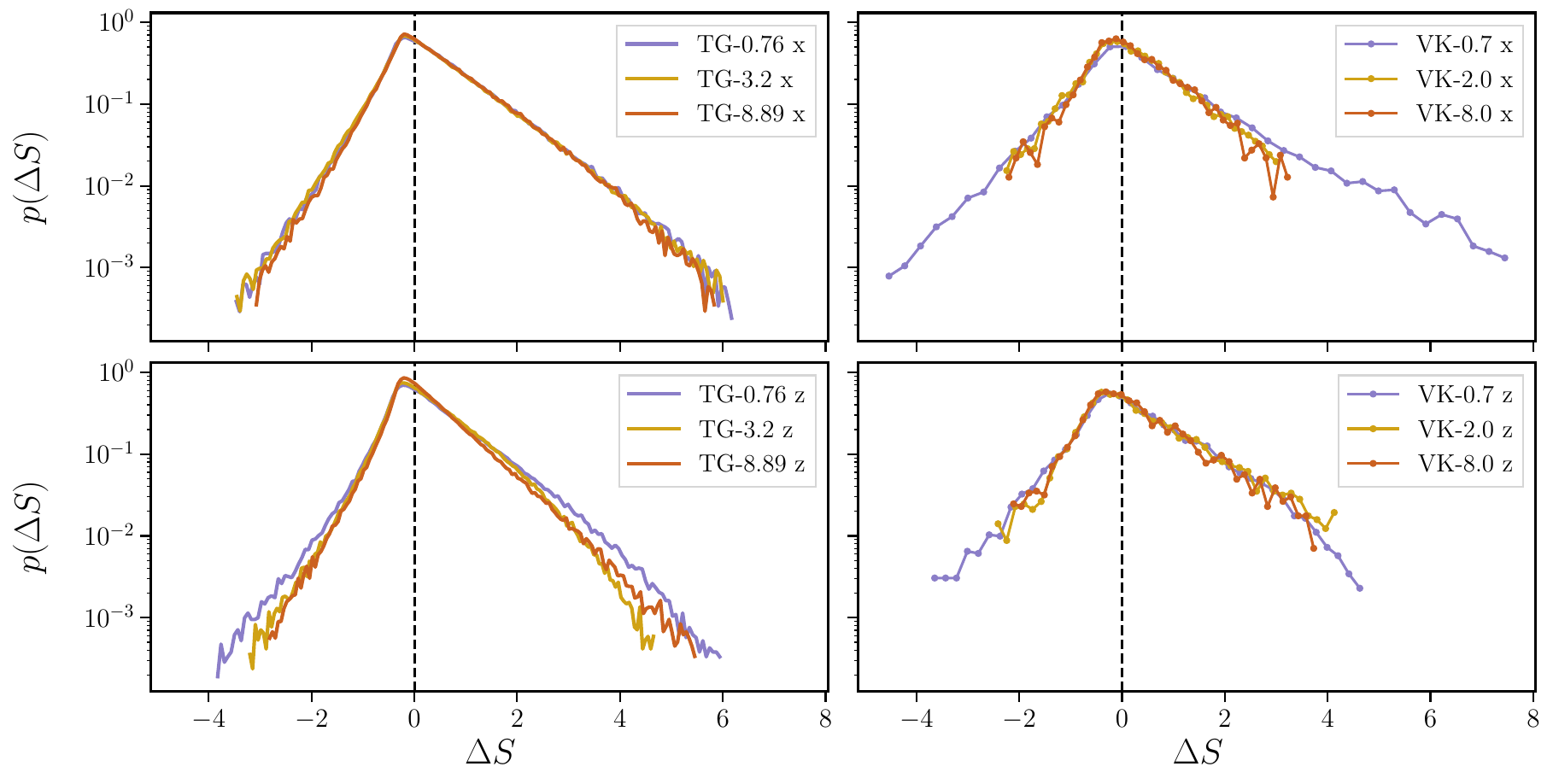}
	\caption{PDFs of total entropy production $\Delta S$ for all datasets. Left panels correspond to simulations, and right panels to experimental data. The top row shows results for the $x$ velocity component, while the bottom row shows the $z$ component.}
	\label{fig:entropies}
\end{figure}

With this stochastic description provided by the scale dependent Fokker--Planck equation, the total entropy variation can be computed individually for each cascade trajectory as the sum of two contributions, $\Delta S = \Delta S_\text{sys} + \Delta S_\text{med}$ \citep{Seifert_2005, Seifert_2012, Nickelsen_2013, Reinke_2018}.
The first term, $\Delta S_\text{sys}$, represents the information entropy change associated with the evolution of the probability distribution from an initial state $p(u_{\tau_i}, \tau_i)$ to a final state $p(u_{\tau_f}, \tau_f)$,
\begin{equation}
	\Delta S_\text{sys} = -\ln\left[\frac{p(u_{\tau_i}, \tau_i)}{p(u_{\tau_f}, \tau_f)}\right].
\end{equation}
The second term, $\Delta S_\text{med}$, quantifies the entropy exchanged with the medium as the system evolves from scale $\tau_i$ to $\tau_f$, and it captures the irreversibility of the cascade trajectories:
\begin{equation}
	\Delta S_\text{med} = \int_{\tau_i}^{\tau_f}
	\left[\partial_\tau u_\tau \, \frac{D^{(1)} - \partial_{u_\tau} D^{(2)} / 2}{D^{(2)}}\right]
	\, \dd \tau.
\end{equation}
Figure \ref{fig:entropies} shows the probability density function (PDF) of entropy production $\Delta S$ computed over all cascade trajectories, for all datasets and for both the $x$ and $z$ velocity components.

\begin{figure}
	\centering
	\includegraphics[width=\linewidth]{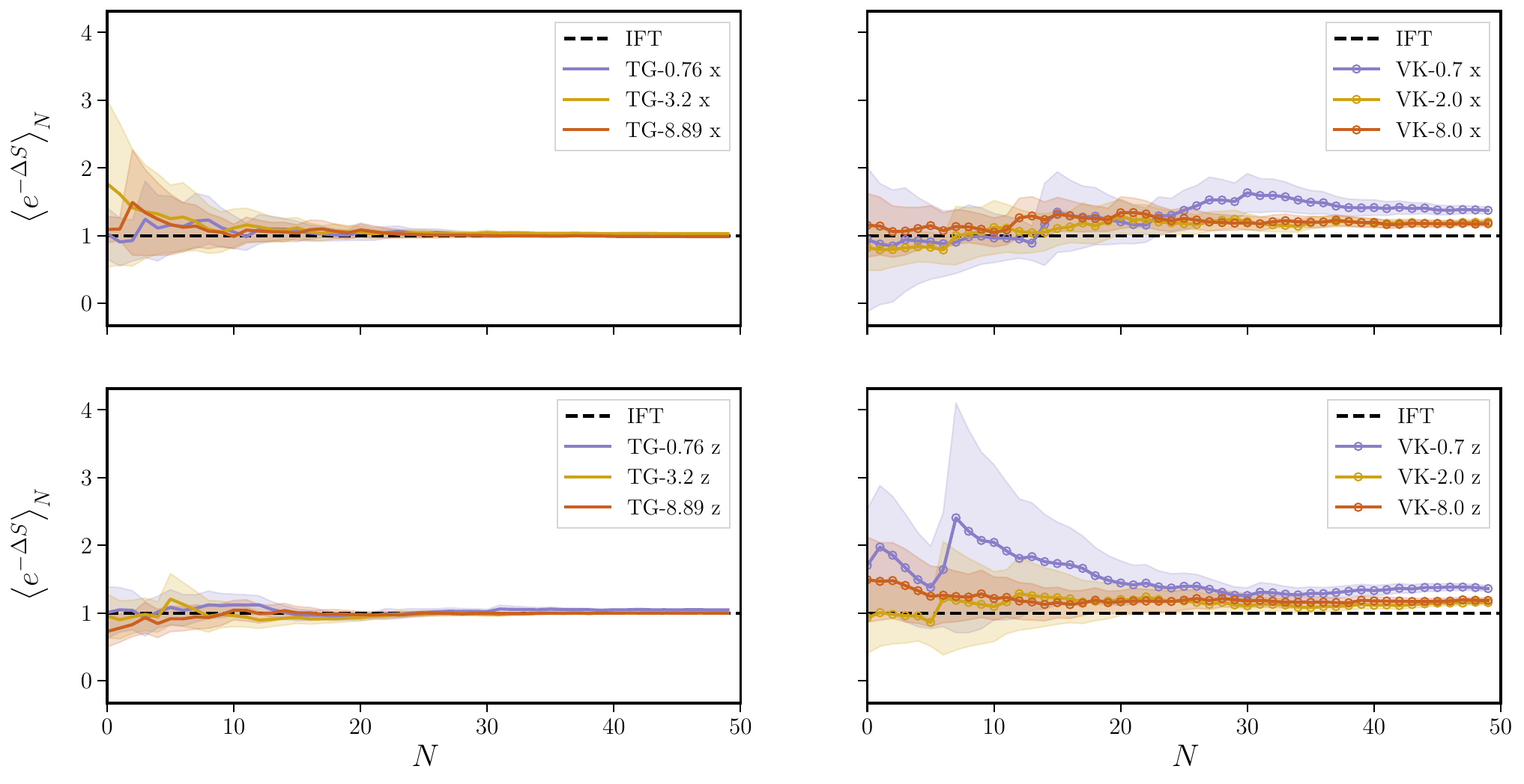}
	\caption{Average of $e^{-\Delta S}$ as a function of the number of cascade trajectories $N$. Left panels correspond to simulations, and right panels to experiments. The top row shows results for the $x$ velocity component, while the bottom row corresponds to the $z$ component. The black dashed line indicates the theoretical prediction from the IFT. Shaded bands represent the 95\% confidence intervals estimated from statistical sampling.}
	\label{fig:ift}
\end{figure}

For all datasets it is more likely to obtain positive entropy increments than negative ones, consistent with the irreversible nature of turbulence. Negative entropy production events are associated with the inverse cascade, and thus with the formation of large-scale structures, while positive entropy production corresponds to the direct cascade \citep{Fuchs_2022}. From this perspective, the net entropy balance reflects the predominance of the direct cascade, as expected for typical three-dimensional turbulence.
In particular, we expect the distribution of entropy variations to satisfy the integral fluctuation theorem:
\begin{equation} \label{eq:ift}
	\ev{e^{-\Delta S}}_N = \int_N e^{-\Delta S}\,p(\Delta S)\, \dd \Delta S = 1,
\end{equation}
which is a fundamental law of non-equilibrium stochastic thermodynamics \citep{Seifert_2012}.
Figure~\ref{fig:ift} shows the ensemble average of $e^{-\Delta S}$ as a function of the number of randomly selected cascade trajectories. All datasets yield values close to $1$, which is indicated by the black dashed line, when a sufficiently large number of trajectories is considered, in agreement with Eq.~\eqref{eq:ift}.
This result is nontrivial, as this quantity is highly sensitive to the specific structure of the drift and diffusion coefficients. Due to the exponential form of the average, even small errors in these coefficients can propagate and amplify rapidly, potentially causing the estimate to diverge \citep{Renner_2001}.
For this reason, both the functional form of the coefficients $D^{(1,2)}(u_\tau, \tau)$ and the agreement with the IFT support the validity of a Fokker--Planck description of particles in the inertial range, even in the presence of strong particle inertia and anisotropic flow conditions.
This result is also consistent with the simplified Markov property verified at the beginning of this study.

\begin{figure}
	\centering
	\includegraphics[width=\linewidth]{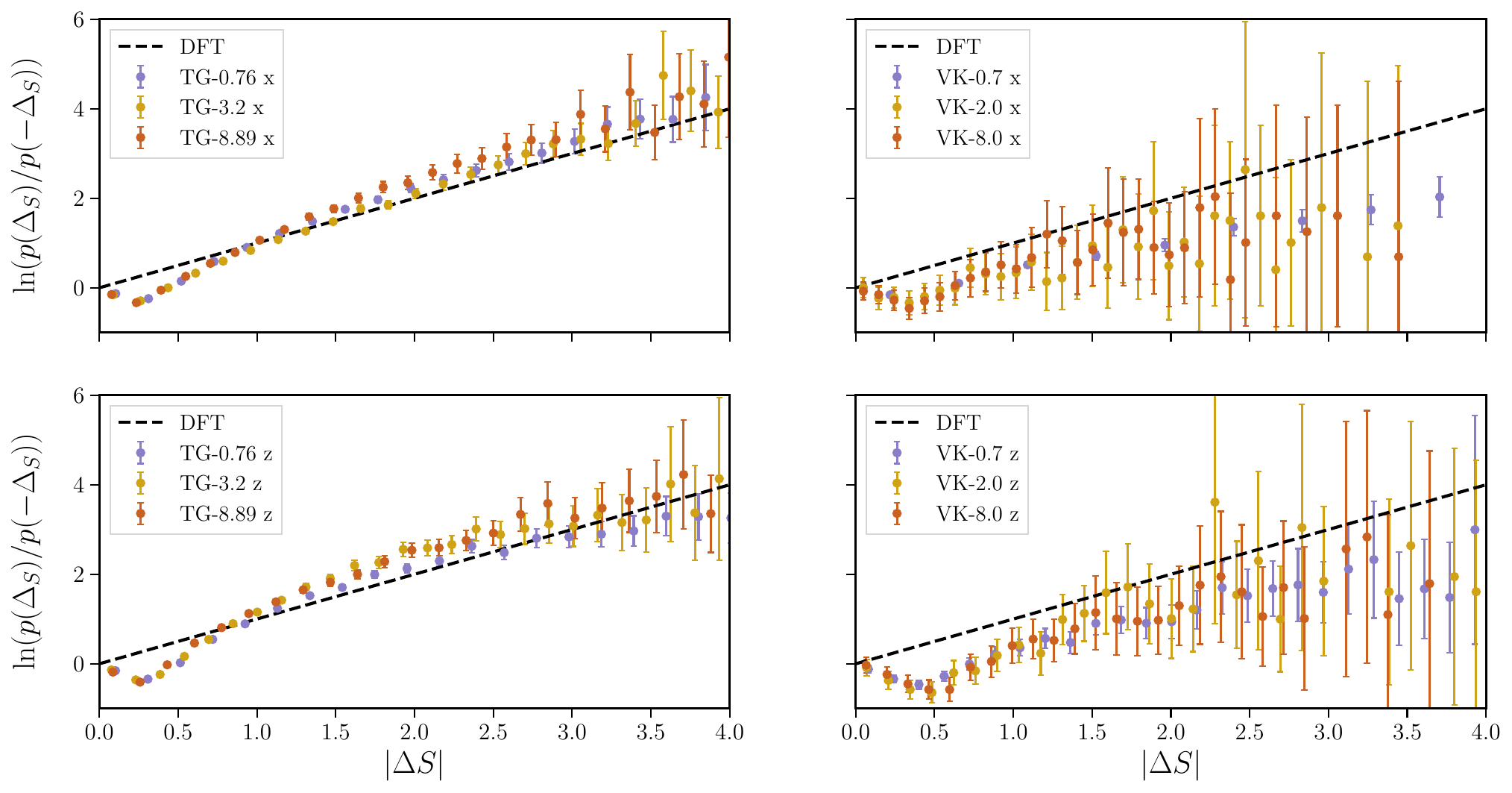}
	\caption{Test of the DFT. The vertical axis shows the logarithm of the ratio of the probability of observing positive entropy production $\Delta S$ to that of observing a negative fluctuation of the same magnitude, as a function of $|\Delta S|$. Left panels correspond to simulations, and right panels to experiments. The top row shows results for the $x$ velocity component, while the bottom row shows the $z$ component. Error bars represent statistical uncertainties.}
	\label{fig:dft}
\end{figure}

It is also worth examining whether our data satisfies the detailed fluctuation theorem. This theorem states that the probability of observing a cascade trajectory that produces a negative entropy variation $-\Delta S$ is exponentially smaller than the probability of a trajectory that produces a positive variation of the same magnitude, i.e.,
\begin{equation}
	\ln\left[\frac{p(\Delta S)}{p(-\Delta S)}\right] = \Delta S.
\end{equation}
While the IFT must hold for a Fokker--Planck description of trajectory entropy production $\Delta S$ to be valid \citep{Seifert_2012}, the DFT need not.
The IFT is a general consequence of stochastic thermodynamics under quite broad conditions, while the DFT relies on additional assumptions such as time-reversal symmetry or steady-state dynamics, which may not apply to non-stationary settings or to turbulent flows in general.
Figure~\ref{fig:dft} shows $\ln[p(\Delta S) / p(-\Delta S)]$ as a function of $\Delta S$. Agreement with the DFT corresponds to alignment with the black dashed line of slope $1$. All datasets are reasonably consistent with this prediction within their estimated uncertainties. Interestingly, however, a systematic deviation (a ``dip'') appears at small values of $|\Delta S|$ in all laboratory and numerical datasets.

\begin{figure}
	\centering
	\includegraphics[width=\linewidth]{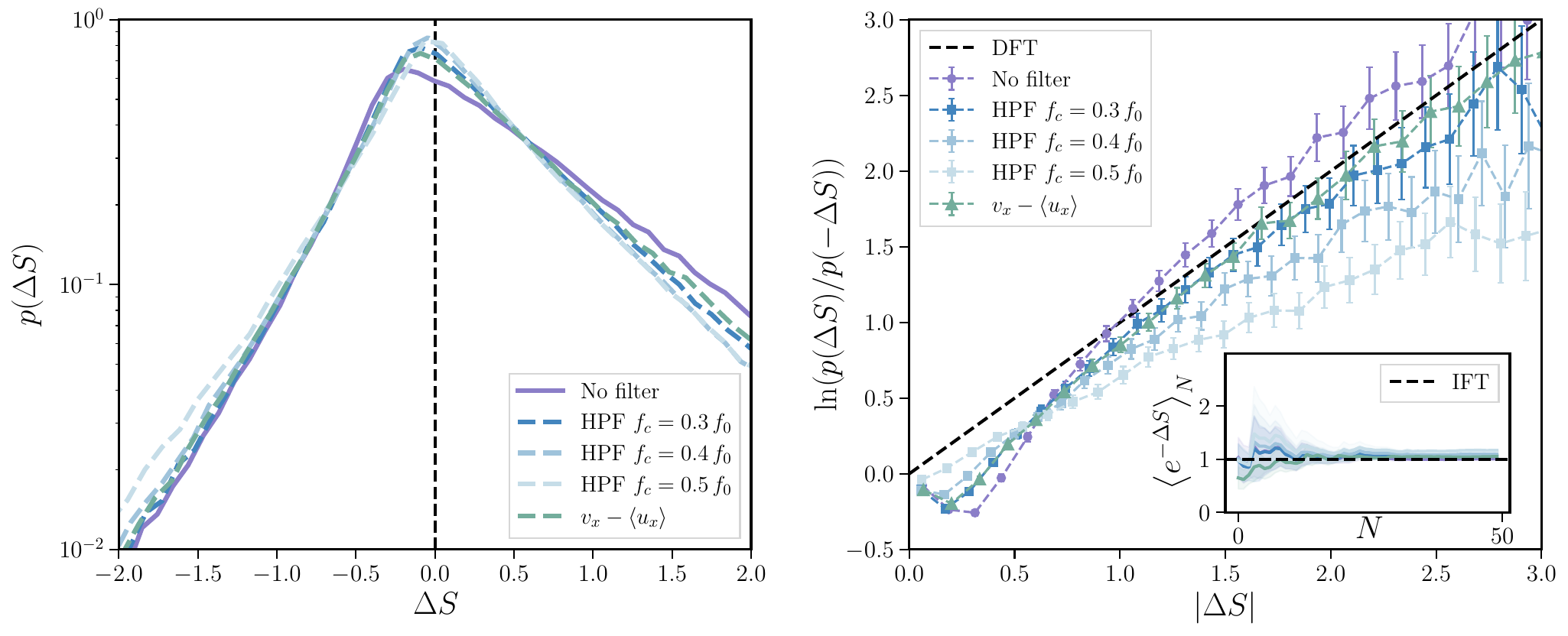}
	\caption{
		PDFs of $\Delta S$ (left) and DFT test (right) for data processed with high-pass filters (HPFs) at different cutoff frequencies for the $x$ velocity component of the TG-0.75 dataset, 
        and for the same component of the particles' velocities, $v_x$, with the Eulerian mean fluid velocity $\langle u_x\rangle$ removed. The inset shows the IFT test for the same processed data. Similar behaviour is observed in all datasets.
	}
	\label{fig:dft_nomean}
\end{figure}

The presence of this dip is noteworthy, as it is less prominent in Lagrangian studies of homogeneous and isotropic turbulence (HIT) \citep{Fuchs_2022}. This suggests that it may be attributed to the influence of the TG and VK mean flows. Such an interpretation is consistent with the thermodynamic framework: the deviation is associated with a systematic shift in entropy production, as suggested by the fact that the mode of $p(\Delta S)$ is not centered at zero. This in turn implies that there is a persistent source of entropy even in the absence of fluctuations, which can be associated with the sweeping of particles by the mean flow. Moreover, the fact that this shift occurs toward negative values of $\Delta S$ suggests the presence of an underlying ordered structure in the system, which is also compatible with the influence of the mean flow. 
To test this hypothesis, we remove the mean flow using two independent procedures, and examine the resulting impact on the entropy PDFs and the DFT analysis. The first method applies a fifth-order Butterworth high-pass filter to the particles' velocities with cutoff frequencies $f_c / f_0 = 0.3$, $0.4$, and $0.5$ (where $f_0 = 1/T$). The second method computes the position-dependent Eulerian mean fluid velocity, and subtracts the mean flow from each particle at each position and time. In both methods, removing the slow components 
shifts the mode of the entropy PDF closer to zero and, correspondingly, improves the agreement with the DFT at small $|\Delta S|$, in line with our hypothesis, while it does not strongly affect the agreement with the IFT. At the same time, the filtering method introduces artificial correlations at small scales, altering the structure of the coefficients $D^{(1)}$ and $D^{(2)}$ when the cutoff frequency gets close to $f_0$, and consequently affecting the statistics of large entropy production events, which are typically associated with intense small-scale fluctuations. This is somewhat expected: removing a mean flow is more complex than simply filtering the signal, and the filter inevitably alters the signal introducing spurious correlations. While additive Gaussian noise in the data has little effect on the results, small systematic correlations can, however, significantly degrade the agreement with the DFT.

\section{Conclusions}

We have shown that velocity increments of inertial particles in turbulent flows, obtained from both DNSs and laboratory experiments with an inhomogeneous mean flow and anisotropy, exhibit Markovian behaviour over the inertial range. This property enables a Fokker--Planck description of particle dynamics, from which drift and diffusion coefficients can be extracted directly from the data.
The inferred coefficients display Stokes and scale-dependent features, consistent with inertial filtering and intermittency, and reveal distinct anisotropic effects in turbulent swirling flows, with both similarities and differences between experiments using neutrally buoyant finite-size particles, and simulations which employ idealized point particle models. Importantly, these differences do not undermine the validity of the Fokker--Planck description. It should be noted that in all cases this stochastic approach provides a deep grasp of the statistics of Lagrangian turbulence and of inertial particles, enabling access to the general joint probability density $p(u_{\tau_j}, u_{\tau_{j+1}}, ..., u_{\tau_{j+n}})$ as different successive scales $\tau$ are considered.

Furthermore, we demonstrated that entropy production along particle cascade trajectories satisfies the IFT with high precision, confirming the quality of the estimated Fokker--Planck equation.
Furthermore, we show consistency with the DFT under appropriate conditions. These results establish a link between stochastic thermodynamics and the statistical description of particle-laden turbulence, opening the way for statistical models of turbulent transport. A perspective raised by this work concerns the effect of the non-ergodic nature of inertial particle trajectories within the present formalism. Indeed, particle clustering and the time irreversibility of turbulent particle trajectories should be somehow reflected in the Kramers--Moyal coefficients and entropy PDFs, a problem left for a future study.

\appendix
\section{Comparison with homogeneous isotropic turbulence}

\begin{figure}
	\centering
	\includegraphics[width=\linewidth]{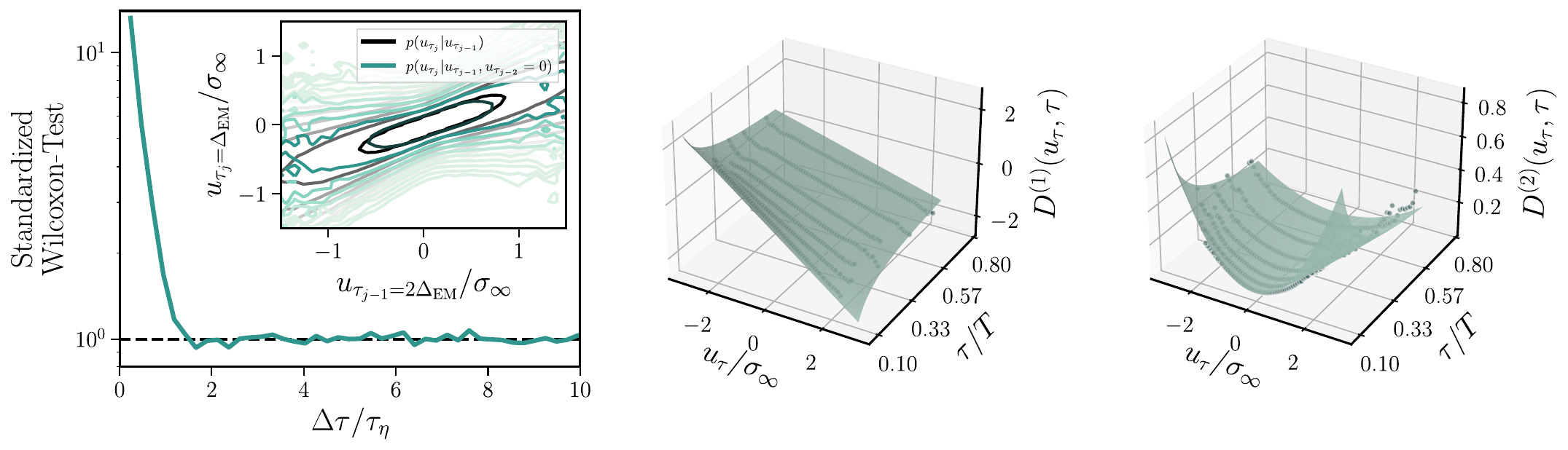}
	\caption{Results for tracers in the JHTDB homogeneous isotropic turbulence simulation with $\text{Re}_\lambda = 433$. The left panel shows the standardized Wilcoxon test statistic as a function of the time lag, with the inset comparing the single-conditioned distribution and the double-conditioned distribution at the same time scales as in Fig.~\ref{fig:2dpdf}. The middle panel shows the drift coefficient $D^{(1)}(u_\tau,\tau)$, while the right panel shows the diffusion coefficient $D^{(2)}(u_\tau,\tau)$.}
	\label{fig:JHTDB_data}
\end{figure}

As a reference, we provide a comparison of the non-equilibrium formulation for tracers in the $1024^3$ pseudospectral simulation of homogeneous and isotropic turbulence with $\text{Re}_\lambda = 433$ from the Johns Hopkins Turbulence Database (JHTDB) \citep{Perlman_2007, Li_2008}. We used $10^5$ tracer trajectories spanning approximately $2.5$ turbulent integral correlation times, and computed velocity increments $u_\tau$ following the same procedure as for the TG and VK datasets.

We tested the Markov property explicitly using single- and double-conditioned PDFs and the Wilcoxon test, and computed the drift and diffusion coefficients $D^{(1)}(u_\tau,\tau)$ and $D^{(2)}(u_\tau,\tau)$. With these coefficients, all other results can be reproduced. The main results are summarized in Fig.~\ref{fig:JHTDB_data}.
The Wilcoxon test indicates that the data is compatible with the Markov property, starting even at smaller values of $\Delta \tau/\tau_\eta$ than the inertial particles considered in this study. This is to be expected as tracers follow the flow perfectly, while inertial particles have their own response time. As a reference, \citet{Fuchs_2022} considered tracers in $512^3$ simulations of homogeneous isotropic turbulence with $\textrm{Re}_\lambda \approx 240$, and found $\Delta_\textrm{EM}/\tau_\eta \approx 7$. The decrease with increasing $\textrm{Re}_\lambda$ is consistent with the physical expectation that a broader inertial range should enhance the local-in-scale character of the cascade \citep{Mininni_2008}. 

The drift and diffusion coefficients obtained from the JHTDB data display the same qualitative dependence as in the TG and VK datasets, with comparable amplitudes at large and intermediate scales $\tau/T$ even when the large-scale flow and the particles' properties are different. The main differences can be seen at the smallest resolved scales. The larger value of $\text{Re}_\lambda$ and the instantaneous response of the tracers result in a broader inertial range, allowing computation of the drift and diffusion coefficients down to time lags $\tau/T \approx 0.1$. At these lags, the trends observed in the Kramers--Moyal coefficients become more pronounced: the slope of $D^{(1)}$ steepens further as $\tau$ decreases, and the curvature of $D^{(2)}$ increases, reflecting stronger scale-dependent relaxation and more intense intermittency as the dissipation range is approached. These results suggest that increasing $\text{Re}_\lambda$ does not substantially modify the qualitative form of the Kramers--Moyal coefficients, but can shift the onset of Markovian behaviour and the intensity of intermittency observed at the smallest scales, in agreement with the usual picture of turbulence in which larger $\text{Re}_\lambda$ results in a broader cascade range.

\backsection[Funding]{PDM and BLE acknowledge financial support from UBACyT Grant No. 20020220300122BA, and from Proyecto REMATE of the Redes Federales de Alto Impacto, Argentina.
The authors acknowledge support of the Johns Hopkins Turbulence Database (JHTDB), which is an open-access platform supported by the National Science Foundation.}

\backsection[Declaration of interests]{The authors report no conflict of interest.}

\backsection[Data availability statement]{The code GHOST used to prepare the $768^3$ simulations is publicly available at \url{https://doi.org/10.5281/zenodo.15472741}. The software used for the statistical analysis is publicly available at \url{https://github.com/bersp/turbfpe}. The data of homogeneous isotropic turbulence on a $1024^3$ grid from the Johns Hopkins Turbulence Database is publicly available at \url{https://doi.org/10.7281/T1KK98XB}.}  

\backsection[Author ORCIDs]{
B.L.~Espa\~nol, \url{https://orcid.org/0000-0001-7957-7344};
M.~Obligado, \url{https://orcid.org/0000-0003-3834-3941};
J.~Peinke, \url{https://orcid.org/0000-0002-0775-7423};
M.~Noseda, \url{https://orcid.org/0000-0003-1864-5408};
P.J.~Cobelli, \url{https://orcid.org/0000-0003-2968-1877};
P.D.~Mininni, \url{https://orcid.org/0000-0001-6858-6755}}

\bibliographystyle{jfm}
\bibliography{bib}

\end{document}